# Quantum Random Number Generation using a Solid-State Single-Photon Source

Simon J. U. White,[1] Friederike Klauck,[2] Toan Trong Tran,[1] Nora Schmitt,[2] Mehran Kianinia,[1] Andrea Steinfurth,[2] Matthias Heinrich,[2] Milos Toth,[1] Alexander Szameit,[2] Igor Aharonovich,[1] and Alexander S. Solntsev[1*]

[1]School of Mathematical and Physical Sciences, Faculty of Science, University of Technology Sydney, Ultimo, NSW, 2007, Australia
[2]Institut für Physik, Universität Rostock, Albert-Einstein-Straße 23, 18059 Rostock, Germany
*Corresponding author: alexander.solntsev@uts.edu.au

(Dated: January 28th 2020)

Quantum random number generation (QRNG) harnesses the intrinsic randomness of quantum mechanical phenomena. Demonstrations of such processes have, however, been limited to probabilistic sources, for instance, spontaneous parametric down-conversion or faint lasers, which cannot be triggered deterministically. Here, we demonstrate QRNG with a quantum emitter in hexagonal boron nitride — an emerging solid-state quantum source that can generate single photons on demand and operates at room temperature. We achieve true random number generation through the measurement of single photons exiting one of four integrated photonic waveguides, and subsequently, verify the randomness of the sequences in accordance with the National Institute of Standards and Technology benchmark suite. Our results open a new avenue to the fabrication of on-chip deterministic random number generators and other solid-state-based quantum-optical devices.

## 1. Introduction

The fundamental unpredictability inherent in genuine random numbers is vital for truly secure encryption [1,2], data science [3], and fundamental research [4,5]. Yet obtaining true randomness turns out to be a highly nontrivial task: Many conventional RNGs are actually pseudo-random, and, at their core, require a trusted source of randomness to expand with deterministic algorithms [6]. While such pseudo-random sequences can be obtained with great speed and efficiency, they tend to be subject to long term correlations. Beyond being a mere nuisance in data science and fundamental research, low-quality random number generators introduce critical points of failure in cryptographic applications [7,8].

A particularly elegant approach to the generation of sequences of fundamentally random numbers are measurements of multipartite quantum states [9]. Here the pursuit of randomness can be simply realized through the measurement of quantum superposition, that is, the measurement of a particle which exists in multiple states simultaneously. In this vein, a wide range of platforms for quantum random number generation (QRNG) have been implemented ranging from the prototypical examples of radioactive decay [10], vacuum fluctuations [11], and laser phase fluctuations [12], to single photons in path superposition [13] and even device-independent realizations [14].

Light, and specifically single photons, is attractive as it offers numerous advantages to quantum information processing, communication, sensing, and miniaturization in optically-integrated chips [15]. In particular, substantial efforts are being undertaken to utilize solid-state sources that can be triggered on-demand, such as quantum dots [16], single molecules [17], or color centers in wide bandgap materials [18]. Of the latter, point defects that act as single-photon emitters (SPEs) in the van der Waals material hexagonal boron nitride (hBN) have recently gained great attention owing to their exceptional brightness [19], photostability [20] and tunability [21]. The applicability of these emitters for quantum communications has also recently been explored with recent demonstrations of Fourier transform limited line widths [22], and the realization of photonic crystal cavities that allow for Purcell enhancement and applications in cavity quantum electrodynamics [23].

On-chip coupling of single-photon emitters, state manipulation, and detection are fundamental for integrated photonics and scalable quantum devices. While the coupling of quantum emission from point defects in 2D materials to optical waveguides has been achieved [24–26], the crucial subsequent step of on-chip photonic quantum state manipulation remains an ongoing challenge towards harnessing hBN emitters for scalable integrated quantum photonics. In the present work, we bridge this gap and make use of a compact and discrete photonic circuit fabricated in fused silica with femtosecond laser direct-writing. The device is designed to efficiently evolve a single photon in a single spatial mode to a multipath superposition state via a simultaneous radial coupling. Until now, all QNRG has been realized without the use of deterministic single-photon sources. Here we address this critical issue and demonstrate QNRG employing defects in hBN, at room temperature, integrated with a compact laser-written photonic chip.

## 2. Experimental Implementation

The hexagonal boron nitride used in our experiments was prepared using commercial multilayer flakes dispersed in solution (Graphene Supermarket). For specific samples to be identified and targeted in a readily reproducible fashion, the hBN solution was drop-cast on a marked silicon substrate and subsequently annealed at 850°C under 1 Torr of argon atmosphere to activate the emitters. The finalized sample allowed us to choose from a wide range of different multilayer hBN flakes with diameters between 100 and 500 nm, and a thickness in the range of 10 - 40 nm [27]. A photoluminescence (PL) map of the sample was taken using a custom-built scanning confocal microscope [28], revealing the location of the desired point-defect SPE in hBN. The chosen defect was excited using a 100 µW, 532 nm laser, and its emission was collected using a 0.9 NA objective (Nikon), as illustrated schematically in Fig. 1(a). The photoluminescence spectrum of the emitter used in our subsequent experiments is shown in Fig. 1(b). To isolate the zero-phonon line (ZPL) from the phonon sideband emission, a 630 ± 20 nm band-pass filter was rotated to center over the ZPL, as highlighted in Fig. 1(b). The excited-state lifetime $\tau_1$ is measured based on the autocorrelation data as $\tau_1 \approx 1.9$ ns, which is comparable with literature values for an hBN SPE [29]. To confirm the single-photon emission from the defect, second-order autocorrelation functions, $g^2(0)$, were measured using a fiber-based Hanbury-Brown-Twiss interferometer. The normalized histogram, seen in Fig. 1(c), is not background-corrected and exhibits a minimum, $g^2(0) = 0.179$, well below the threshold value of 0.5, unequivocally indicating single-photon emission from a single defect.

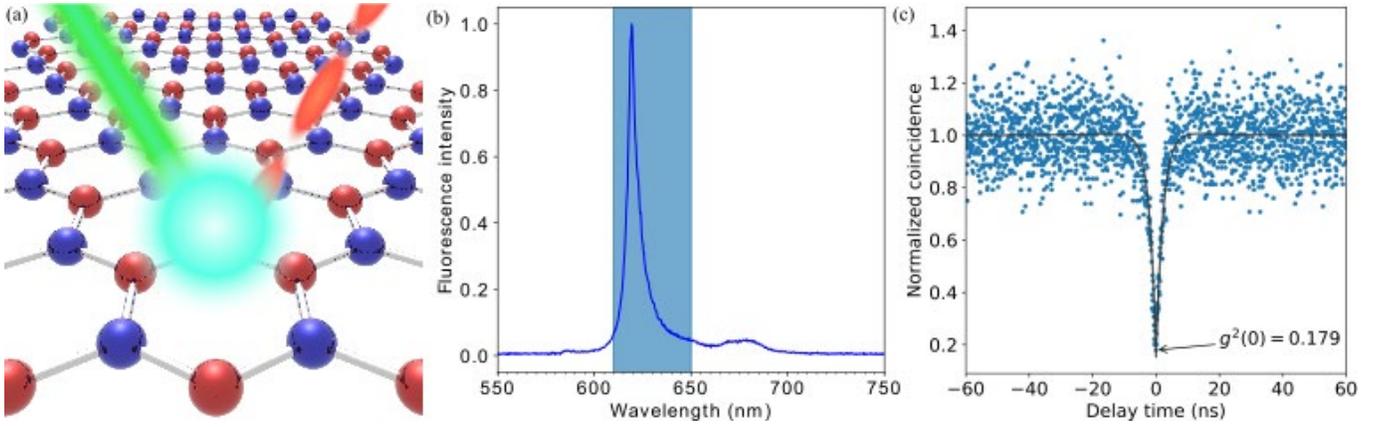

Fig. 2. Hexagonal boron nitride single-photon emitter. (a) A point defect in the hexagonal lattice of boron and nitrogen atoms is excited by a 532 nm laser to emit a stream of discrete photons. (b) Photoluminescence spectrum of the hBN defect, comprised of a sharp zero-phonon line at 619 nm and phonon side-band at ~675 nm. The single-photon emission is filtered using a 630 ± 20 nm band-pass filter, shaded in blue. (c) The antibunching curve of the spectrally filtered emission (blue dots) confirms room temperature single-photon emission, with a minimum $g^2(0)$ value of 0.179.

As scalability and robustness are key features of integrated optical waveguides, we employed a compact splitter arrangement to synthesize spatially superposition states from the single photons injected into its input port. In this entirely passive structure, a ring-shaped arrangement serves to distribute the probability amplitude simultaneously between nine waveguides by means of evanescent coupling (see Fig. 2) [30]. Notably, in contrast to previous chip designs employing cascaded sequences of directional couplers [13], the functional region of this chip is invariant in propagation direction and, therefore, free of radiation losses associated with repeated S-bends. The chip was fabricated in fused silica (Corning 7980) using the femtosecond direct laser writing technique [31]. 150 fs pulses from a commercial laser system (Coherent Mira/RegA) with a 100 kHz repetition rate at a carrier wavelength of 800 nm were focused approximately 200 µm below the sample surface using a standard microscope objective (20x, NA=0.36). By moving the sample with respect to the focal spot with a positioning system (Aerotech Inc.), waveguides with a refractive index contrast of approximately 5×10$^{-4}$ were created along the desired trajectories. At a wavelength of 815 nm, the waveguides feature propagation losses in straight sections are below 0.3 dB/cm and show an elliptical mode profile of 9x13 µm². In order to account for the anisotropic coupling behavior, the active region of the splitter was vertically elongated to an ellipse with half axes of 20 µm and 23 µm, respectively. To maximize the extraction rate, the central guide was inscribed with a slightly lower propagation constant phase-matched to a specific super-mode of the outer ring. Once the desired multipartite spatial superposition state has evolved, a fan-out section serves to separate the individual channels to a planar arrangement with 127µm pitch to ensure compatibility with commercially available fiber arrays.

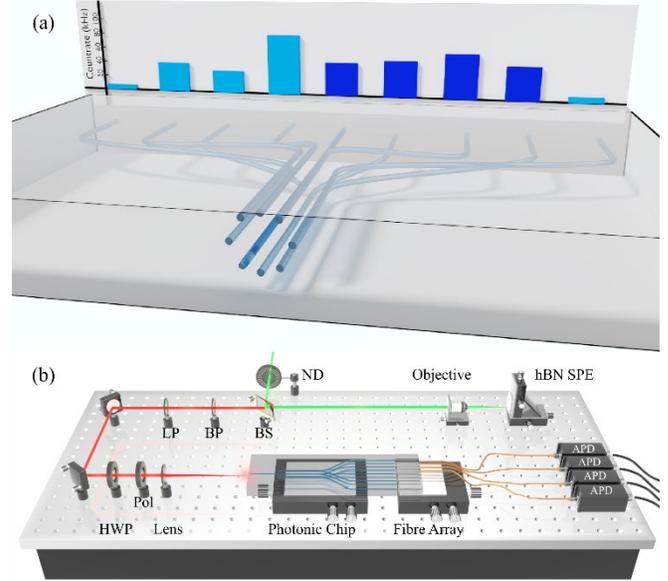

Fig. 2. Schematic of the chip. (a) Light from a single input waveguide is coupled to a ring structure of eight waveguides, which are then bent into an equidistant array for out-coupling. Above the structure, the corresponding output single-photon count-rates are displayed, where the four chosen waveguides (dark blue) have a standard deviation of 11.6 % before induced losses. (b) Schematic of the experimental setup. A hexagonal boron nitride single-photon source is excited with a 300 µw, 532 nm CW laser and photons are collected through a 0.9 NA objective. The laser excitation is suppressed using a 568 nm long pass filter (LP) and the zero-phonon line is selected using a 630 ± 20 nm bandpass filter (BP). The polarization is then controlled using a half wave plate (HWP) and a linear polarizer (Pol). Light is directly coupled to the chip using a lens and collected using a multimode fiber array. Photons are detected with avalanche photodiodes (APD).

# 3. Results and Discussion

The hBN single-photon emission is directly coupled to the photonic chip through free space with a lens, as illustrated in Fig. 2. The emitter is excited to produce single photons at a rate of ~1 MHz before the chip; after the chip across all nine waveguides, we achieved total throughput of ~350 kHz. Here, losses are mainly due to coupling to and from the chip as well as bending losses within the fan-out section of the waveguides. Single photons at the output of each waveguide are coupled to a multimode fiber array and detected using avalanche photodiodes (Excelitas) with efficiencies of 65%, dark counts < 100 Hz, and dead times of 22 ns. Their respective arrival times were recorded using a time tagger with 10 ps resolution (Swabian Instruments). To verify that the single-photon emissions of the hBN source were efficiently coupled to the chip and additional background counts were negligible, we recorded second-order correlations between pairs of output channels ($g^2_{1,5}(0)$ for channels 1 and 5). With minimal losses, the single-photon purity is maintained, and is unequivocally demonstrated with fitted $g^2(0)$ values of < 0.24, as seen in Fig. 3(a).

To obtain a random binary sequence, we record time-tagged photon arrivals across four avalanche photodiodes (APDs). To maximize the generation rate, we employ a scheme where one photon represents two bits by interpreting each detection event as a two-digit binary number, i.e. '00', '01', '10', or '11', which are recorded in succession. Notably, this approach can be naturally scaled up: With the creation of larger multiplexed states distributed across $2^n$ outputs, a single photon represents $n$ bits, i.e., $2^n$ numbers, each with probabilities of $2^{-n}$. Due to the quadratic scaling of resources, we chose four of APDs for our proof-of-principle experiments, corresponding to the four channels highlighted in dark blue, seen in Fig. 2(a). The single-photon count rates after the chip for channels 1 – 4 were 45, 49, 60, and 45 kHz, respectively. To ensure precise equal probability amplitudes on each detector, we introduced coupling losses of up to ~25%, until each channel received 45 kHz upon detection. The amplitude distribution, defined as a function of average count rate, was 25.0%, 25.0%, 25.1%, and 24.9% in channels 1 – 4, respectively. Notably, the outstanding brightness of our hBN emitter yielded a total post-chip count rate of 189 kHz, despite collecting photons from only four of the nine channels.

The randomness of the data stream was evaluated using a python implementation of the National Institutes of Standards and Technologies (NIST) test suite for random and pseudorandom number generators [32,33]. The test suite determines the probability (p-value) that the generated sequence is a random sequence. The tests were performed on sequences of 1,000,000 bits, and for a sequence to be considered random, the p-value of each test must exceed 0.01, given a significance level of α = 0.01. Each test has a different interpretation summarized by its shorthand name. For example, the Frequency tests determine the absolute ratios of 0's and 1's per section (Block) or the entirety (Monobit) of the sequence. The Runs tests determine whether uninterrupted sequences of identical bits occur as expected for a random sequence. Template tests identify the probability of specific short sequences appearing throughout the generated sequence. And the Discrete Fourier Transform (Spectral) Test detects periodic signatures within the sequence that would indicate the sequence is non-random. Further information on the rest of the tests and the implementation is available on the NIST website [32]. Successfully passing all tests, as summarized in Fig. 3(b) and Supplement 1, we demonstrate the first solid-state single-photon source coupled with a static multiplexing chip for use in random number generation applications. Crucially, certain mismatches of count rates between different detectors may be inevitable in most real-world scenarios, e.g., due to fabrication limitations, spectrally broad emission lines which affect well-defined coupling constants, as well as physical differences in the individual detectors. Yet, given the superior brightness of solid-state single-photon sources, these can be easily remedied by selective introduction of appropriate losses in individual channels, and further mitigated by the use of narrow-bandwidth hBN emitters [34,35].

Finally, we also considered the generation of four simultaneous streams of random numbers from a single source, demonstrating the scalability of multiplexing given the limitations of current photodetector technology. With dead times of 22 ns, the single-photon avalanche diodes used have a maximum theoretical flux around 45 MHz. Multiplexing effectively reduces the relative flux using $n$ independent channels. To generate the random binary sequence, each channel is independently analyzed, taking the integer photon arrival time in picoseconds modulo 2. The NIST test suite was once more applied, and all four independent channels separately pass the required significance (p-value > 0.01) for all tests (see Supplementary Fig. S2). Here we realize two random number generation schemes from a single static device coupled with bright room-temperature single-photon sources. The flexibility of the spatial multiplexing is not limited to RNG applications but provides a platform for the exploration of quantum cryptography and quantum information processing [36–38].

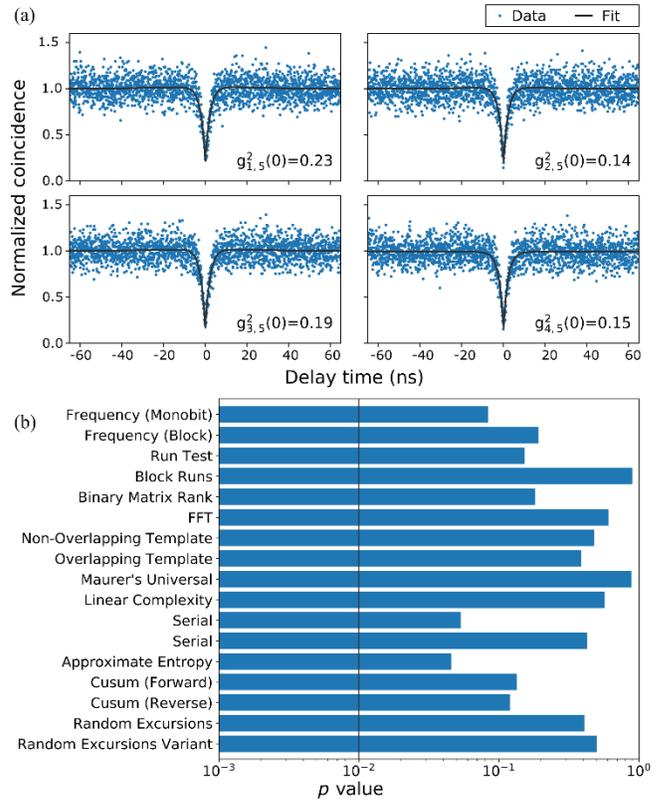

Fig. 3. Single-photon correlation and random number statistics of the multiplexed state. (a) Antibunching curves show that single-photon purity is retained throughout the chip. Each output is correlated with output 5 and the second order correlation minimum is always significantly below 0.5. (b) NIST randomness test results for a sequence of 1M bits, generated using the arrival of single photons in four spatial modes. The black vertical line displays the minimum p-value required to indicate true randomness (p-value > 0.01). Test parameters are available in the supplementary information.

## 4. Concluding Remarks

In this work, we demonstrate on-chip quantum state manipulation and quantum random number generation using photons produced by a room-temperature single-photon solid-state emitter. We use the hexagonal boron nitride platform due to its high single-photon brightness and show efficient coupling to an on-chip photonic waveguide structure, allowing us to prepare a spatial superposition state and generate a stream of true quantum random numbers. The photonic chip utilized in this experiment can be readily integrated with standard optical fibers and other photonic devices. Together with the exceptional brightness and robustness of hBN-based single-photon sources, the system presented here offers an intrinsically stable and readily scalable platform for photonic quantum information processing.

**Funding.** Australian Research Council (DE180100070, DP190101058); Universities Australia and Deutscher Akademischer Austauschdienst (DAAD); Deutsche Forschungsgemeinschaft (SZ 276/12-1, BL 574/13-1, SZ 276/15-1, SZ 276/20-1); Alfried Krupp von Bohlen and Halbach Foundation; European Union's Horizon 2020 research and innovation program (800942).

**Acknowledgments.** The authors would like to thank L. Maestrini and T. Biesenthal for valuable discussions on statistics and C. Otto for preparing the high-quality fused silica samples for the inscription of all photonic chips used in our experiments.

**Disclosures.** The authors declare no conflicts of interest

See Supplement 1 for supporting content.